\documentclass[epj,referee]{svjour}
\usepackage{amssymb}
\usepackage{latexsym}
\usepackage{graphics}
\begin{document}
\title{
Deconfined criticality for the two-dimensional
quantum $S=1$-spin model with the three-spin and biquadratic interactions
}
% \subtitle{Do you have a subtitle?\\ If so, write it here}
\author{Yoshihiro Nishiyama  %First author\inst{1} \and Second author\inst{2}% etc
}                     % Do not remove
\offprints{}          % Insert a name or remove this line
\institute{Department of Physics, Faculty of Science,
Okayama University, Okayama 700-8530, Japan}
\date{Received: date / Revised version: date}
\abstract{
The criticality between the nematic and
valence-bond-solid (VBS)
phases was investigated 
for
the two-dimensional quantum $S=1$-spin model with
the three-spin and
biquadratic interactions
by means of the numerical diagonalization method.
It is expected that the criticality belongs to 
a novel universality class,
the so-called deconfined criticality,
accompanied with unconventional critical indices.
In this paper,
we incorporate the three-spin interaction, and 
adjust the (redundant) interaction parameter
so as to optimize
the finite-size behavior.
Treating the finite-size cluster with $N \le 20 $ spins,
we estimate the correlation-length
critical exponent as $\nu=0.88 (3)$.
\PACS{
{75.10.Jm}        {Quantized spin models} \and
{05.30.-d}        {Quantum statistical mechanics} \and
{75.40.Mg} {Numerical simulation studies} \and
{74.25.Ha}        {Magnetic properties}
     } % end of PACS codes
} %end of abstract
\maketitle
%

%Sandvik
%   staggered na dimer  : long range dimer interaction promotes the 1st order pt

\section{\label{section1}Introduction}

The phase transition between the N\'eel and valence-bond-solid (VBS)
phases for the two-dimensional quantum spin system
is attracting much attention recently 
\cite{Senthil04,Senthil04b,Levin04,Sandvik07,Melko08,Kuklov04,Kuklov08,Jian08,Kruger06,Chen13,Kotov09,Isaev10b,%
Kaul11,Kaul12,Harada13,Pujari13}; 
see Ref. \cite{Kaul13} for a review.
It is expected that the phase transition,
the so-called deconfined criticality,
belongs to a novel universality class,
accompanied with unconventional critical indices.
Originally,
the idea was developed \cite{Senthil04,Senthil04b,Levin04} in the context of the 
gauge-field-theoretical description for the
two-dimensional strongly-correlated-electron system.
Meanwhile, it turned out that the underlying physics is common
to a variety of systems in terms of the emergent gauge field 
\cite{Hikami79,Nogueira12,Wang14,Toldin14}.

%CPN を積極的にひくこと
  % hikami Grover

As a lattice realization of the deconfined criticality,
the quantum $S=1/2$-spin square-lattice antiferromagnet with the plaquette four-spin interaction,
the so-called $J$-$Q$ model \cite{Sandvik07,Melko08},
has been investigated extensively;
the
bipartite-lattice systems such as the square- 
\cite{Sandvik07,Melko08}
and honeycomb-lattice \cite{Harada13,Pujari13}
antiferromagnets  
do not
conflict with the quantum Monte Carlo method,
and large-scale-simulation results are available.
However,
it is still unclear whether the phase transition is
critical \cite{Sandvik07,Melko08,Isaev10b,Kaul11,Kaul12,Harada13,Pujari13}
%%%%%%%%%%%%%%%%%%%%%%%%%
%%%%%%%%%%%%%%%%%%%%%%%%%
%%%%%%%%%%%%%%%%%%%%%%%%%
%%%%%%%%%%%%%%%%%%%%%%%%%
%%%%%%%%%%%%%%%%%%%%%%%%%
%%%%%%%%%%%%%%%%%%%%%%%%%
%%%%%%%%%%%%%%%%%%%%%%%%%
%%%%%%%%%%%%%%%%%%%%%%%%%
%%%%%%%%%%%%%%%%%%%%%%%%%
% or a weak first-order transition with an appreciable latent-heat release
 or a weak first-order transition with a latent-heat release
\cite{Kuklov04,Kuklov08,Jian08,Kruger06,Chen13,Kotov09}. 
The controversy may be reconciled by
a recent renormalization-group analysis \cite{Bartosch13},
which revealed 
an influence of a notorious
marginal operator around the deconfined-critical fixed point;
see Ref. \cite{Sandvik10} as well.
Because the character of the marginal operator depends on  
each lattice
realization, it may be sensible to survey a variety of
lattice realizations.

\begin{figure}
\includegraphics{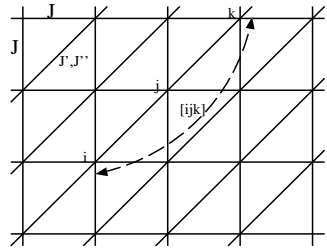}%
\caption{  \label{figure1}
The square-lattice model with the skew-diagonal interaction is considered;
see the Hamiltonian (\ref{Hamiltonian}).
The nearest-neighbor interaction $J$ 
stabilizes the nematic phase \cite{Harada02},
whereas the skew-diagonal biquadratic and three-spin interactions, $J'$ and $J''$,
respectively,
lead to a formation of dimers \cite{Fath95,Michaud12};
the phase diagram is presented in Fig. \ref{figure2}.
the symbol $[ijk]$ stands for the adjacent three sites along
the skew-diagonal chain.
%In the pioneering study \cite{Harada07},
%the interaction $J'$ (spatial anisotropy)  
}
\end{figure}

The $S=1$-spin model is a clue to the realization of the deconfined
criticality \cite{Harada07,Grover11,Nishiyama11}.
A key ingredient is that the $S=1$-spin model 
admits 
the biquadratic interaction, which stabilizes the VBS phase 
as the spatial anisotropy varies \cite{Fath95}.
The phase transition separating the VBS and nematic 
\cite{Harada02} phases
is expected to belong to the deconfined criticality \cite{Grover11}.
We consider a non-bipartite-lattice version 
(Fig. \ref{figure1});
the details and underlying ideas are explained afterward.
In the preceding paper \cite{Nishiyama11},
the 
correlation-length critical exponent was estimated as $\nu=0.92(10)$.
In this paper, based on the preceding studies \cite{Harada07,Grover11,Nishiyama11},
we incorporate a rather novel interaction term intrinsic to the $S=1$-spin model,
namely, the three-spin-interaction term \cite{Michaud12,Wang13} (in addition to the
biquadratic one), and survey the extended parameter space so as to 
optimize the finite-size behavior.

To be specific,
we present the Hamiltonian
for the two-dimensional $S=1$-spin model
\begin{eqnarray}
{\cal H} &=&
- J \sum_{\langle ij \rangle} 
          [j {\bf S}_i \cdot {\bf S}_j + ({\bf S}_i\cdot {\bf S}_j)^2 ]
-J' \sum_{\langle \langle ij \rangle \rangle} ({\bf S}_i\cdot{\bf S}_j)^2
\nonumber \\
\label{Hamiltonian}
& &
+J'' \{ \sum_{\langle \langle ij \rangle \rangle} {\bf S}_i\cdot{\bf S}_j   \nonumber \\
& &
       +
j_3 \sum_{[ijk]} [({\bf S}_i \cdot {\bf S}_j)({\bf S}_j \cdot {\bf S}_k)
+h.c.]
     \}
   .
\end{eqnarray}
Here, the symbol
${\bf S}_i$ denotes the quantum $S=1$-spin operator placed
at each square-lattice point $i$
(Fig. \ref{figure1}).
The summations,
$\sum_{\langle ij \rangle}$,
$\sum_{\langle \langle ij \rangle \rangle}$,
and
$\sum_{[ijk]}$,
run over all possible rectangular (nearest-neighbor) edges $\langle ij\rangle$,
skew-diagonal pairs $\langle \langle ij \rangle \rangle$, and 
skew-diagonal adjacent three sites $[ijk]$, respectively.
Correspondingly, the coupling constants,
 $J$, $J'$, and $J''$,
denote the nearest-neighbor-, skew-diagonal- and 
skew-diagonal-adjacent-three-spin-interaction parameters.
Hereafter, the coupling constant $J'$ is considered as the unit of energy ($J'=1$).
The underlying physics behind each interaction parameter is as follows.
For sufficiently large nearest-neighbor interaction $J(>0)$, the system reduces
to a two-dimensional model,
and the nematic phase emerges \cite{Harada02}; 
here,
the quadratic component of the Heisenberg interaction is set to
$j=0.5$ throughout this study.
On the contrary, the coupling constants $J'(=1)$ \cite{Fath95}
and $J''(>0)$ \cite{Michaud12,Wang13}
strengthen the spatial
anisotropy, promoting the formation of singlet dimers along the
skew-diagonal bonds; a schematic phase diagram is presented in Fig. \ref{figure2}.
%Actually, as for $J=0$,
%the phase transition separating the nematic and VBS
%phases was observed numerically; the phase diagram is presented in Fig. .
In this paper, we incorporate the latter interaction 
term
and adjust this (redundant)
interaction
parameter $J''$ as well as the three-spin-interaction component $j_3$ 
so as to optimize the finite-size-scaling behavior.

It has to be mentioned that in the pioneering study \cite{Harada07},
the bipartite-lattice version (without the diagonal interaction) of 
Eq. (\ref{Hamiltonian}) was investigated 
by means of the quantum Monte Carlo method;
for the bipartite lattice,
the spatial anisotropy $J'$ inevitably
violates the symmetry between the horizontal and vertical directions,
and the asymmetry 
might alter the nature of the
transition \cite{Harada13}.
Our non-bipartite-lattice version (\ref{Hamiltonian}) retains the
symmetry (between the horizontal and vertical directions),
as would be apparent from Fig. \ref{figure1}.
In order to cope with the non-bipartite-type lattice,
 we employ the exact diagonalization method with the aid
of Novotny's method (screw-boundary condition)
\cite{Novotny90},
which enables us to treat a variety of system sizes
$N=10,12,\dots$ in a systematic manner;
note that the number of spins constituting a rectangular cluster
is restricted within $N=4,9,\dots$.

The rest of this paper is organized as follows.
In Sec. \ref{section2}, we present the simulation results.
The technical details as to the screw-boundary condition are presented as well.
In Sec. \ref{section3}, we address the summary and discussions,
providing a brief overview on the past studies
of the
correlation-length critical exponent $\nu$.

\section{\label{section2} Numerical results}

In this section,
we present the simulation results.
To begin with, we explain
the simulation scheme to implement the screw-boundary 
condition, namely,
Novotny's method
\cite{Novotny90}, briefly.
Owing to this scheme, we are able to treat an arbitrary number of
spins, $N=10,12,\dots$,
 constituting a two-dimensional cluster.
The linear dimension $L$ of the cluster is given by
$L=\sqrt{N}$, because the $N$ spins form a rectangular cluster.

\begin{figure}
\includegraphics{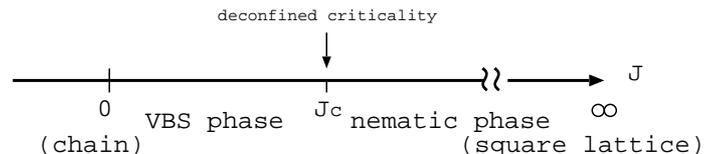}%
\caption{  \label{figure2}
A schematic phase diagram for the Hamiltonian 
(\ref{Hamiltonian})
is presented.
As the interaction $J(/J')$ increases,
 the VBS \cite{Fath95,Michaud12} and nematic 
\cite{Harada02} phases appear successively.
Typically, the transition point locates around $J_c \approx 0.3$
for small $J''(/J')$.
The phase transition point,
namely, the deconfined criticality \cite{Harada07,Grover11},
is our concern.
}
\end{figure}

%%%%%%%%%%%%%%
%%%%%%%%%%%%%%
%%%%%%%%%%%%%%
The Hamiltonian (\ref{Hamiltonian})
has been investigated extensively in some limiting cases.
In order to elucidate 
the phase diagram, Fig. \ref{figure2},
we recollect a number of related studies \cite{Fath95,Harada02,Michaud12,Wang13};
we also address a brief account of the
parameter range
surveyed
in our preliminary study.
In the limit $J\to\infty$, the model reduces to the two-dimensional Heisenberg model
with the biquadratic interaction. 
According to Ref. \cite{Harada02},
around $j=0.5$, the nematic phase is realized.
With $J=0$ and $J''=0$, the one-dimensional biquadratic-interaction Heisenberg model is realized,
and the VBS phase emerges \cite{Fath95}.
Similarly, with $J=0$ and $J'=0$, 
the VBS phase is realized for sufficiently large $j_3> 0.111$ \cite{Michaud12}.
Hence, the interaction parameter $J$ 
interpolates smoothly these limiting cases,
and the phase diagram, Fig. \ref{figure2}, follows. 
In the preliminary stage of the research,
we dwelt on the subspace $J'' = 0$, which was studied in Ref. \cite{Nishiyama11}.
Turning on the interaction $J''$ gradually, we arrive at the optimal regime $J'' \approx 0.08$,
as indicated in Fig \ref{figure5}.
%%%%%%%%%%%%%%%%%%%
%%%%%%%%%%%%%%%%%%%
%%%%%%%%%%%%%%%%%%%
At least for a moderate range of $J''<0.2$, a fundamental feature of the phase diagram,
Fig. \ref{figure2}, is kept maintained;
namely, no signature such as an appearance of a certain intermediate phase could
be detected.

\subsection{\label{section2_1}Simulation method: Screw-boundary condition}

In this section, we explain
Novotny's method \cite{Novotny90}
to implement
the screw-boundary condition.
The screw-boundary condition
enables us to treat a variety of
system sizes $N=10,12,\dots$;
note that naively,
the system size is restricted within  quadratic numbers 
$N=4,9,\dots$
for a rectangular cluster.

In this paper,
we follow the simulation scheme reported
in Ref. \cite{Nishiyama11}, where the $J''$ term 
of the Hamiltonian
(\ref{Hamiltonian}) was not taken into account.
The missing term is incorporated by the 
addition
of the following term to Eq. (5) of Ref. \cite{Nishiyama11};
\begin{equation}
\label{Jdd_term}
J''\sum_{i=1}^N \{ {\bf S}_i\cdot{\bf S}_{i+1}  
+j_3[({\bf S}_i\cdot{\bf S}_{i+1})({\bf S}_{i+1}\cdot{\bf S}_{i+2}) +h.c.] \}.
\end{equation}
(The index $i$ runs over a one-dimensional alignment
$i=1,2,\dots,N$ in a way intrinsic to the screw-boundary condition.)
Thereby,
we diagonalize the Hamiltonian matrix given by Eq. (5) of
Ref. \cite{Nishiyama11} with the term (\ref{Jdd_term}),
employing
 the Lanczos algorithm
for a finite-size cluster
with $N \le 20$ spins.
Rather technically, 
the diagonalization was performed within
the zero-momentum subspace,
at which
the magnetic- (triplet-) excitation gap $\Delta E$ opens.

\subsection{\label{section2_2}
Finite-size scaling of $\Delta E$: Critical point}

In this section, based on the simulation method explained in
Sec. \ref{section2_1},
we evaluate
the excitation gap 
$\Delta E$.
Thereby, we estimate
the location of the critical point 
via the scaling analysis of $\Delta E$.

\begin{figure}
\includegraphics{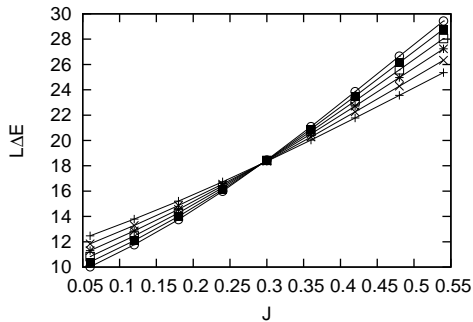}%
\caption{  \label{figure3}
The
scaled energy gap $L \Delta E$
is plotted for various $J(/J')$ and the system sizes of
($+$) $N(=L^2)=10$,
($\times$) $12$,
($*$) $14$,
($\Box$) $16$,
($\blacksquare$) $18$,
and
($\circ$) $20$;
here, the interaction parameters were
set to $J''(/J')=0.08$ and $j_3 = 0.45$.
The intersection point of the curves,
$J_c \approx 0.3$, indicates the location of the critical point.
}
\end{figure}

In Fig. \ref{figure3},
we plot the scaled energy gap $L \Delta E$ for various $J(/J')$ and
$N=10,12,\dots,20$
($L=\sqrt{N}$);
here, the interaction parameters are set to $J''(/J')=0.08$ and
$j_3=0.45$.
According to the finite-size-scaling theory,
the intersection point 
of $L \Delta E$
indicates a location of the critical point $J_c \approx 0.3$,
because the scaled energy gap should be scale-invariant
(dimensionless) at the critical
point.

\begin{figure}
\includegraphics{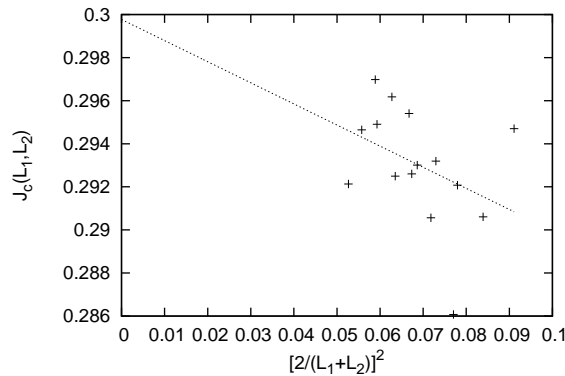}%
\caption{  \label{figure4}
The approximate critical point 
$J_c(L_1,L_2)$
(\ref{approximate_critical_point})
is plotted 
for $[2/(L_1+L_2)]^2$
($10 \le N_1 < N_2 \le 20$).
Here, the interaction parameters 
are the same as those of Fig. \ref{figure3}.
The least-squares fit to the data yields
$J_c=0.2998(45)$ in the thermodynamic limit.
An oscillatory finite-size deviation 
(series of bumps around $N\approx9,16,\dots$) is
due to the screw-boundary condition \cite{Novotny90}.
}
\end{figure}

In Fig. \ref{figure4}, we plot 
the approximate critical point $J_c(L_1,L_2)$ for $[2/(L_1+L_2)]^2$
with $10 \le N_1 <N_2 \le 20$ ($L_{1,2}=\sqrt{N_{1,2}}$);
the interaction parameters are the same as those of Fig. \ref{figure3}.
Here,
the
approximate critical point $J_c(L_1,L_2)$
is
defined by the formula
\begin{equation}
\label{approximate_critical_point}
L_1 \Delta E(L_1)|_{J=J_c(L_1,L_2)}=
L_2 \Delta E(L_2)|_{J=J_c(L_1,L_2)}    ,
\end{equation}
for a pair of system sizes $(L_1,L_2)$.
The least-squares fit to the data 
in Fig. \ref{figure4}
yields $J_c=0.2998(45)$ in the thermodynamic limit
$L \to \infty$.
This extrapolated value does not affect the subsequent analysis
(Sec. \ref{section2_3}),
and we do not go into the discussion of the extrapolation error;
actually, 
the approximate critical point $J_c(L_1,L_2)$ (rather than the extrapolated $J_c$)
is fed into the formula, Eq. (\ref{approximate_critical_exponent}).
% shusoku no yosa

Last, we address a number of remarks.
First, in Fig. \ref{figure4},
the finite-size behavior 
seems to be oscillatory;
actually, the data exhibit successive bumps for quadratic values of
$N(=L^2)\approx 9,16,\dots$.
Such an oscillatory deviation is an artifact 
\cite{Novotny90}
of the screw-boundary condition.
Second,
the set of 
the coupling constants $J''=0.08$ and $j_3=0.45$ 
were determined 
so as to optimize the finite-size behavior of Fig. 
\ref{figure4} (as well as Fig. \ref{figure5} mentioned afterward).
Actually,
as for $J'' \ne 0.08$ and $j_3 \ne 0.45$, 
the finite-size behavior of $J_c(L_1,L_2)$
suffers from steep finite-size drift
and enhanced
bumps.
%To suppress such a finite-size drift,
%we adjust the coupling
%constants to the above mentioned values.
% details as to Fig. \ref{figure5} are explained in the next section.
%%%%%%%%%%%%%%%
%%%%%%%%%%%%%%%
%%%%%%%%%%%%%%%
%%%%%%%%%%%%%%%
Last, in the scaling analysis, Fig. \ref{figure3}, we 
postulated the dynamical critical exponent $z=1$.
Here, we followed the conclusion
$z=1$ obtained in the pioneering studies
\cite{Sandvik07,Melko08}.

\subsection{\label{section2_3}Correlation-length critical exponent $\nu$}

\begin{figure}
%%\resizebox{0.4\textwidth}{!}{%
\includegraphics{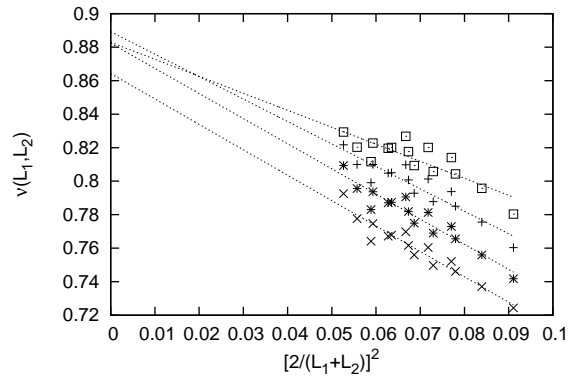}%
%%}
\caption{  \label{figure5}
the approximate critical exponent 
$\nu(L_1,L_2)$ (\ref{approximate_critical_exponent})
is plotted for $[2/(L_1+L_2)]^2$ ($10\le N_1 < N_2 \le 20$)
with the interaction parameters,
($+$) $(J'',j_3)=(0.08,0.45)$,
($\times$) $(0,0.45)$,
($*$) $(0.04,0.45)$,
and
($\Box$) $(0.12,0.45)$.
The least-squares fit to these data yields
$\nu=0.889(10)$, 
$0.8642(87)$,
$0.8823(93)$
and 
$0.883(12)$,
respectively, in the thermodynamic limit.
Clearly, 
in an optimal regime $0.04 \le J'' \le 0.12$,
the extrapolation can be taken reliably;
see text for details.
An oscillatory finite-size deviation 
(bump around $N\approx16$) is
an artifact of
the screw-boundary condition \cite{Novotny90}.
%the bump grows, as the parameter $J''$ increases.
}
\end{figure}

In this section, we estimate the correlation-length critical exponent $\nu$.
Based on the approximate critical point
$J_c(L_1,L_2)$ (\ref{approximate_critical_point}),
we are able to calculate the approximate correlation-length critical exponent
\begin{equation}
\label{approximate_critical_exponent}
\nu(L_1,L_2)
=\frac
{\ln (L_1/L_2)
} 
{\ln \{ \partial_J [L_1\Delta E(L_1)]/ \partial_J [L_2\Delta E(L_2)] \} |_{J=J_c(L_1,L_2)}  }
 ,
\end{equation}
for a pair of system sizes $(L_1,L_2)$.
In Fig. \ref{figure5}, as the symbol $+$,  
we plot $\nu(L_1,L_2)$ for  $[2/(L_1 + L_2)]^2$ with $10\le N_1<N_2 \le 20$
($L_{1,2}=\sqrt{N_{1,2}}$); here, the interaction parameters are the same as
those of Fig. \ref{figure3}.
The data exhibit an oscillatory deviation (bump at $N \approx 16$),
which is an artifact of the screw-boundary condition \cite{Novotny90}.
The least-squares fit to the data in Fig. \ref{figure5}
yields $\nu=0.889(10)$ in the thermodynamic limit.

Similar analyses were carried out independently 
for various values of $J''$ with $j_3=0.45$ fixed.
As a consequence,
the approximate critical point is plotted
in Fig. \ref{figure5}
 for
($\times$) $J''=0$,
($*$) $0.04$,
and
($\Box$) $0.12$.
The least-squares fit to these data yields
 $\nu=0.8642(87)$,
$0.8823(93)$,
 and $0.883(12)$, respectively,
in the thermodynamic limit.
Notably enough, the 
interaction parameter $J''$ governs
the convergence of $\nu(L_1,L_2)$ to the thermodynamic limit.
Particularly,
in the optimal regime
$0.04 \le J'' \le 0.12$,
the extrapolation to the thermodynamic limit
can be taken reliability,
allowing us to appreciate the critical exponent 
$\nu \approx 0.88 $
rather systematically.
For exceedingly large $J''>0.12$,
there emerge a notable finite-size drift (to the counter direction)
as well as enhanced oscillatory deviations (steep bumps around $N \approx 16$)
as to $\nu(L_1,L_2)$ and even $J_c(L_1,L_2)$,
preventing us from extrapolating $\nu$
unambiguously.
Surveying various $J''$ as well as $j_3$,
we observed that the uncertainty of the extrapolation
is bounded by $\Delta \nu=0.03$.
%The series of results indicate $\nu\approx$.
%As mentioned above, the oscillatory deviation with the amplitude $..$
%may be an indicator of the extrapolation error;
%actually, the deviation amplitude $kk$ dominates the least-squares-fit error
%$ll$.
As a result, we estimate the correlation-length critical exponent as
\begin{equation}
\label{critical_exponent}
\nu=0.88(3)
  .
\end{equation}
The result is consistent with the estimate 
$\nu=0.92(10)$ reported in Ref. \cite{Nishiyama11},
 where the three-spin
interaction
was not taken into account.

%%%%%%%%%%%%%%%%%%%
%%%%%%%%%%%%%%%%%%%
%%%%%%%%%%%%%%%%%%%
%%%%%%%%%%%%%%%%%%%
%%%%%%%%%%%%%%%%%%%
%%%%%%%%%%%%%%%%%%%
%%%%%%%%%%%%%%%%%%%
In Fig. \ref{figure5},
we assumed implicitly that
the dominant scaling correction 
should obey the power law $1/L^2$.
However, as mentioned in Introduction,
there might be a logarithmic correction \cite{Bartosch13,Sandvik10},
which is not negligible for large system sizes.
We cannot exclude a possibility that
such a correction gives rise to an ambiguity as to
the estimate (\ref{critical_exponent}).
Here, aiming to provide a crosscheck, we carry out an alternative analysis of 
criticality
as presented in the next section.

\subsection{\label{section2_4}Scaling plot for $\Delta E$}

In this section, we present the scaling plot for $\Delta E$
as 
a 
cross-check of the analyses in Sec. \ref{section2_2} and \ref{section2_3}.

\begin{figure}
\includegraphics{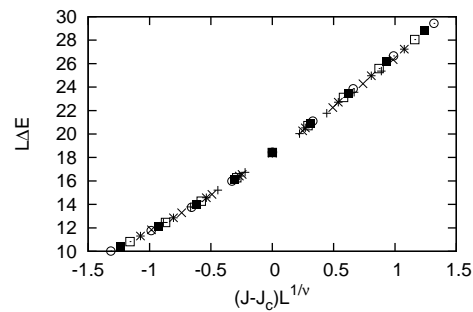}%
\caption{  \label{figure6}
The scaling plot, 
$(J-J_c)L^{1/\nu}$-$L\Delta E$, 
is shown for
($+$) $N(=L^2)=10$,
($\times$) $12$,
($*$) $14$,
($\Box$) $16$,
($\blacksquare$) $18$,
and
($\circ$) $20$;
the interaction parameters are the same as those of Fig.
\ref{figure3}.
Here, the scaling parameters are set to 
$J_c=0.2998 $ and $\nu= 0.88$.
The data collapse into a scaling curve
satisfactorily, providing a cross-check for
the analyses in
Sec. \ref{section2_2} and \ref{section2_3}.
}
\end{figure}

In Fig. \ref{figure6},
we present the scaling plot, 
$(J-J_c)L^{1/\nu}$-$L\Delta E$
for the same interaction parameters as those of Fig. \ref{figure3}.
Here, the scaling parameters are set to
$J_c=0.2998$ (Sec. \ref{section2_2}) and 
$\nu=0.88$ (Sec. \ref{section2_3}).
The scaled data appear to fall into a scaling curve
satisfactorily
for an appreciable range of $J$,
confirming the validity of the analyses in Sec. \ref{section2_2} and \ref{section2_3}.
As mentioned in Introduction,
corrections-to-scaling behavior \cite{Bartosch13,Sandvik10}
may
depend on each lattice realization.
In this respect, we stress that
the present $S=1$-spin model is less affected by severe scaling corrections.
%albeit the expression for the Hamiltonian (\ref{Hamiltonian})
%might be rather complicated.

This is a good position to
address a remark on the validity of the deconfined-criticality scenario.
As mentioned in Introduction, it is still unclear whether the 
(N\'eel-VBS) phase transition
is continuous or not.
The scaling analysis in Fig. \ref{figure6}
suggests that the (spatial-anisotropy-driven nematic-VBS) 
phase transition would be continuous.
A notable point is that the estimate
$\nu=0.88(3)$ 
%%%%%%%%%%%%
%%%%%%%%%%%%
%%%%%%%%%%%%
%%%%%%%%%%%%
%%%%%%%%%%%%
%(\ref{critical_exponent}) is substantially  larger than, for instance, 
(\ref{critical_exponent}) takes a rather enhanced value;
actually, it appears to be larger than, for instance, 
that of 
the Heisenberg universality class,
$\nu=0.7112(5)$ \cite{Compostrini02}.
Because of the following reason,
such a feature might exclude a possibility of the discontinuous
phase transition.
%%%%
%%%%
%%%%
%%%%
%%%%
%%%%
As a matter of fact, 
the discontinuous phase transition
does exhibit a pseudo-critical behavior
(for finite system sizes)
%For the discontinuous phase transition,
%the exponent $\nu$ should take a much smaller value;
%actually, the discontinuous transition does exhibit a pseudo-critical behavior
with an enhanced effective
specific-heat critical exponent  $\alpha=d\nu$ 
($d$: spatial
dimension)
\cite{Borgs90}.
Resorting to the hyper-scaling relation
$\alpha=2- 3 \nu$, one arrives at a suppressed $\nu=0.4$,
which is inaccordant with ours.
%
%which in turn indicates a suppressed $\nu$ via the hyper-scaling relation
%$\alpha=2-d\nu$.
Such a tendency is reasonable, because the discontinuous transition is
accompanied with a latent-heat release, enhancing the 
specific-heat critical exponent $\alpha$ to a considerable extent.
A comparison with other existing estimates for $\nu$
is presented in the next section.

%In general  % C. Borgs R. Kotechk\'y J Stat Phys 61 (1990) 79
%the discontinuous exhibits a quasi-critical behavior with
%an effective critical exponent $alpha/nu(=d)=2$ 
%($\alpha$: specific heat exponent);
%namely, the sudden latent heat release 
%enhances the specific-heat singularity significantly.
%Hence,
%according to 
%the hyper scaling relation, the exponent should be a small value.
%On the contrary, our estimate $nu$ exclude a possibility of (weak) first-order
%transition;
%note that the estimate is even larger than that of ....
%A comparison with existing results is presented in the next section.

\section{\label{section3}Summary and discussions}

The deconfined criticality between
the nematic and VBS phases
for the two-dimensional $S=1$-spin model with both three-spin and biquadratic interactions
(\ref{Hamiltonian})
was investigated with the numerical diagonalization method
 for a finite-size cluster with $N \le 20$ spins;
so far, the case without the three-spin interaction has been investigated
extensively 
\cite{Harada07,Nishiyama11}.
The extended interaction-parameter space
enables us to search for a regime of suppressed finite-size corrections.
Actually, the interaction parameter $J''$ governs
the convergence to the thermodynamic limit,
as shown in Fig. \ref{figure5}.
In fact,
for $0.04 \le J'' \le 0.12$, the extrapolation can be
taken reliably.
As a result,
we estimate the
correlation-length critical exponent
as $\nu=0.88(3)$; the result agrees with the 
preceeding estimate $\nu=0.92(10)$ \cite{Nishiyama11}.

As a comparison, we recollect related studies of the deconfined
criticality,
placing an emphasis on the critical exponent $\nu$.
First,
as mentioned in Introduction, the $J$-$Q$
model has been investigated extensively.
Pioneering studies reported 
   $\nu=0.78(3)$ \cite{Sandvik07}                % Sandvik prl 98 07 227202
and
   $0.68(4)$ \cite{Melko08}.               % melko prl 100 08 017203
Possibly, there emerges the logarithmic corrections
to scaling
in the vicinity of the deconfined criticality 
\cite{Bartosch13,Sandvik10}, preventing us
from identifying
the character of the phase transition definitely.
A recent simulation result for the honeycomb-lattice model 
indicated a rather suppressed 
value of   $\nu=0.54(5)$ \cite{Pujari13}. %  prl 111  13 087203
%It is not unclear whether the transition is discontinuous or not;
%particularly, a recent simulation result reveals a possible deviation
%of the lattice ....
%However, the deconfined criticality intrinsically suffers from the
%logarithmic corrections .....
Second,
a unique approach to the deconfined criticality 
was made by the extention of the internal symmetry 
to SU($N$) with the continuously variable $N$ \cite{Beach09}.
There was reported a clear evidence of the phase transition of
a continuous character
with
the enhanced critical exponent
   $\nu=0.75$-$1$. %  beach prb 80 09 184401
%the conclusion is reminiscent of ours, Eq. (\ref{critical_exponent}).
Third, the ``fermionic'' criticality \cite{Hikami79,Nogueira12} indicates 
$\nu=0.80(3)$ \cite{Wang14} and $\approx 0.88$ \cite{Toldin14}.
%Again, these estimates
%the $1/N$-expansion analysis of a fermion model
%indicates a equivalence to the CP model (deconfined criticality),
%and the fermionic criticality has a relevance to ours.
%A notable point raised by these recent developments is an influence of the
%(almost) marginal operator, which sets a severe obstacle as to the analysis of 
%the criticality concerned.
Last, we overview simulation results for the classical counterparts.
For the dimer model,
the critical exponent was estimated as
    $\nu=0.73(5)$ \cite{Charrier08}, % gauge-field theory picture of  % charr prl 101 08 167205
          $0.5$ \cite{Misguich08}, %         misguich  prb 78 08 100402
         $0.5$-$0.53$ \cite{Charrier10}, and  %       charrier prb 82 2010 014429
           $0.5$-$0.6$ \cite{Papanikolaou10}.  %     papanikolaou prl 104 10  045701
The 
hedgehog-suppressed O($3$) model  
indicates an enhanced exponent
  $\nu=1.0(2)$ \cite{Motruich04}.  % prb 70 04 075104
%%To conclude, 
These conclusions have not been settled yet. 
Nonetheless,
our result indicates a tendency toward an enhancement of $\nu$ such as the fermionic
criticality \cite{Wang14,Toldin14}.

The present simulation result supports the deconfined-criticality scenario
at least for the 
(spatial-anisotropy-driven)
nematic-VBS phase transition;
see the discussion in Sec. \ref{section2_4} as well.
In  Ref. \cite{Michaud13},
a novel two-dimensional quantum-spin model with the three-spin-interaction was introduced.
The model may be a good clue to the realization of the deconfined criticality
without the spatial anisotropy.
However,
its rich phase diagram has not been fully clarified.
A progress toward this direction is remained for the future study.

%namely, the $J$-$Q$ model,
%the
%estimates, $\nu=0.78(3)$ \cite{Sandvik07} and 
%$\nu=0.68(4)$ \cite{Melko08}, were reported.
%The SU($N$) model with the continuously variable $N$ yields \cite{Beach09}
%a similar conclusion $0.75 < \nu < 1$.
%These results do not contradict ours.
%On the one hand, as for the honeycomb-lattice antiferromagnet \cite{Pujari13},
%the result $\nu=0.54(5)$ was obtained.

%%\begin{acknowledgments}
\section*{Acknowledgment}
This work was supported by a Grant-in-Aid 
for Scientific Research (C)
from Japan Society for the Promotion of Science
% from Monbu-Kagakusho, Japan
(Contact No. 25400402).
%%\end{acknowledgments}

%\begin{figure}
%\resizebox{0.75\textwidth}{!}{%
%  \includegraphics{leer.eps}
%\vspace{5cm}       % Give the correct figure height in cm
%\end{figure}

%\begin{figure*}
%\vspace*{5cm}       % Give the correct figure height in cm
%\end{figure*}
%\begin{table}
%\begin{tabular}{lll}
%\hline\noalign{\smallskip}
%\end{tabular}
% Or use
%\vspace*{5cm}  % with the correct table height
%\end{table}
%
% BibTeX users please use
% \bibliographystyle{}
% \bibliography{}

\end{document}